\documentclass[aps,prd,showpacs,thmsa,a4paper,onecolumn]{revtex4}
\begin{document}

\author{Luiz Carlos Ryff}
\title{\textbf{Bell's Theorem Reexamined}}
\affiliation{\textit{Universidade Federal do Rio de Janeiro,
Instituto de F\'{\i}sica,
Caixa Postal 68528, 21945-970 Rio de Janeiro, RJ, Brazil}\\
e-mail: ryff@if.ufrj.br}

\begin{abstract}
An Einstein-Podolsky-Rosen (EPR)-like argument using events
separated by a time-like interval strongly suggestes that measuring
the polarization state of a photon of an entangled pair changes the
polarization state of the other distant photon. Trough a very simple
demonstration, the Wigner-D'Espagnat inequality is used to show that
in order to prove Bell's theorem neither the assumption that there
is a well-defined space of complete states $\lambda $ of the
particle pair and well-defined probability distribution $\rho
(\lambda )$ over this space nor the use of counterfactuals is
necessary. These results reinforce the viewpoint that quantum
mechanics implicitly presupposes some sort of nonlocal connection
between the particles of an entangled pair. As will become evident
from our discussion, relinquishing realism and/or free will cannot
solve this apparent puzzle.
\bigskip\

\end{abstract}
\maketitle

\textbf{I. INTRODUCTION}

\bigskip\ \ As emphasized by Schr\"{o}dinger, entanglement is \textit{the }%
characteristic trait of quantum mechanics.$^{\mathbf{1}}$ Einstein,
Podolsky, and Rosen (EPR) used entangled states to try to prove that this
theory is incomplete,$^{\mathbf{2}}$ and Bell made things clearer by showing
that no local hidden variable theory can mimic quantum mechanics.$^{\mathbf{3%
}}$ Apparently, measuring the state of a particle can instantaneously change
the state of another particle that can be arbitrarily distant from the
first. But no superluminal telegraph can be devised using this phenomenon.
The correlations become evident only when the results, gathered at two
different spatial regions, are compared. In principle, one observer cannot
know what kind of experiment the other is performing. That is, there is no
detectable contradiction with special relativity. Bell's theorem has been
extended to real situations,$^{\mathbf{4}}$ proofs have been introduced that
do not rely on inequalities,$^{\mathbf{5}}$ long-distance experimental tests
of entanglement have been performed,$^{\mathbf{6}}$ and the use of entangled
particles for cryptographic purposes has been proposed.$^{\mathbf{7}}$
However, the mystery remains,$^{\mathbf{8}}$ and even conflicting points of
view on the conceptual significance of Bell's theorem have been presented.$^{%
\mathbf{9}}$ Here I will advocate the viewpoint that quantum mechanics is an
intrinsically nonlocal theory, that is, it implicitly presupposes that
measuring the state of a distant particle of an entangled pair can
instantaneously affect the other's state.

\bigskip\

\textbf{II. THE WEIRDNESS OF QUANTUM ENTANGLEMENT}

\bigskip Let us consider the following situation: a source S emits pairs of
entangled photons, $\nu _1$ and $\nu _2$, in the state
\[
\left| \psi \right\rangle =\frac 1{\sqrt{2}}\left( \left| a,\mathbf{k}%
\right\rangle \left| a,-\mathbf{k}\right\rangle +\left| a_{\perp },\mathbf{k}%
\right\rangle \left| a_{\perp },-\mathbf{k}\right\rangle \right)
\]
\[
=\frac 1{\sqrt{2}}\left( \left| b,\mathbf{k}\right\rangle \left| b,-\mathbf{k%
}\right\rangle +\left| b_{\perp },\mathbf{k}\right\rangle \left| b_{\perp },-%
\mathbf{k}\right\rangle \right)
\]
\begin{equation}
=\frac 1{\sqrt{2}}\left( \left| c,\mathbf{k}\right\rangle \left| c,-\mathbf{k%
}\right\rangle +\left| c_{\perp },\mathbf{k}\right\rangle \left| c_{\perp },-%
\mathbf{k}\right\rangle \right) =...,  \label{1}
\end{equation}
where $\left| a,\mathbf{k}\right\rangle $ $\left( \left| a_{\perp },-\mathbf{%
k}\right\rangle \right) $ represents a photon with polarization parallel
(perpendicular) to $\mathbf{a}$ following direction $\mathbf{k}$ $\left( -%
\mathbf{k}\right) $, and so on.

From $(1)$ we see that quantum mechanical formalism does not allow us to
assign definite polarization states to the photons. This can only be done
when a polarization measurement is performed. For example, if photon $\nu _1$
($\nu _2$) is detected in state $\left| a\right\rangle $, then we
immediately know that the other photon of the pair, $\nu _2$ ($\nu _1$), has
been ``forced'' into the same state $\left| a\right\rangle $. We also see
that $(1)$ does not allow us to predict the outcome of the
measurement---state $\mid a\rangle $ or $\mid a_{\perp }\rangle $, for
example. This strongly suggests that quantum mechanics implicitly
presupposes some sort of superluminal---actually, infinite-speed---
interaction: measuring the state of one of the two particles instantaneously
changes the state of the other. According to Bell, ``in these EPR
experiments there is the suggestion that behind the scenes something is
going faster than light,'' and Bohm declared: ``I would be quite ready to
relinquish locality; I think it is an arbitrary assumption.''$^{\mathbf{10}}$
An important point to be emphasized is that the correlations become evident
when one reads the results that have been automatically registered. The
observer's consciousness does not seem to play any role in the
process.\thinspace \thinspace If this were not so, mind states of distant
observers would have to communicate to reproduce the quantum correlations.

Although the discussion has been centered on the nonlocal aspects of quantum
mechanics, the important question is knowing if a measurement performed on
one of the photons of an entangled pair can change the state of the other.
To examine this problem, it is preferable to consider time-like events. Let
us imagine that the path followed by $\nu _2$ is modified so that the first
observer, Alice (A), after measuring the state of $\nu _1$, can inform the
second observer, Bob (B), about her result before he detects $\nu _2$.
Naturally, assuming that the detection of $\nu _1$ cannot change the
polarization state of $\nu _2$, it is irrelevant whether we consider
space-like or time-like events. On the other hand, if there is some sort of
nonlocal connection between the photons, it may be more illuminating to
discuss situations in which there is no doubt about which one has been
detected first. An aleatory sequence of photons in states $\left|
a\right\rangle $ and $\left| a_{\perp }\right\rangle $ is indistinguishable
from another aleatory sequence of photons in states $\left| b\right\rangle $
and $\left| b_{\perp }\right\rangle $, or $\left| c\right\rangle $ and $%
\left| c_{\perp }\right\rangle $, and so on. Therefore, the detection of
photons $\nu _2$ provides us no information about the orientation of the
polarizer on which photons $\nu _1$ are impinging. That is, without using a
classical communication channel, A cannot use entangled states to send
information to B. But time-like events allow us to try to clarify the
following question: What does it really mean when we say---in agreement with
$(1)$---that measuring the state of the first photon forces the second into
a well-defined polarization state? When we are dealing with space-like
events, it is not possible to assign an objective and well-defined
polarization state to $\nu _2$ just before it impinges on the polarizer,
since the question ``Which photon was \textit{really} detected first?'' is
meaningless. On the other hand, in the case of time-like events, A can send
a message informing about the state of $\nu _2$, and B can then perform a
measurement to check if the information is correct. It seems that EPR's
criterion---``\textit{If, without in any way disturbing a system, we can
predict with certainty (i.e., with probability equal to unity) the value of
a physical quantity, then there exists an element of physical reality
corresponding to this physical quantity.''}$^{\mathbf{2}}$---is perfectly
valid here. Therefore, depending on the orientation of the polarizer on
which $\nu _1$ impinges, $\nu _2$ will reach its polarizer in two possible
states: $\left| a\right\rangle $ and $\left| a_{\perp }\right\rangle $, or $%
\left| b\right\rangle $ and $\left| b_{\perp }\right\rangle $, or $\left|
c\right\rangle $ and $\left| c_{\perp }\right\rangle $, etc. However,
assuming that the detection of $\nu _1$ has no influence on the polarization
state of $\nu _2$, there is no reason for $\nu _2$ to be found only in the
states $\left| a\right\rangle $ and $\left| a_{\perp }\right\rangle $, for
instance. On the other hand, if there is an influence, it still must be
present when space-like events are considered, since the very same
correlations can be observed. From this point of view, there must be a
nonlocal connection between $\nu _1$ and $\nu _2$, unless we assume that a
photon can somehow be in different polarization states simultaneously. We
can also consider the following reasoning: If the outcome of the first
measurement is aleatory, but that of the second becomes well determined
(after we know the result of the first), then, considering the symmetry of
the situation, either (a) the first measurement was not aleatory or (b)
there must be some connection between $\nu _1$ and $\nu _2$. In other words,
if quantum mechanics is complete, it must be nonlocal.

\bigskip\

\textbf{III. BELL'S\ THEOREM}

\bigskip\ The above notions, however, clash with the spirit of relativity,
and we may conjecture that some still unknown local theory exists that
perfectly mimics quantum theory. Bell's theorem shows that this cannot be
so. As a consequence, it seems that our conceptions of space and time must
be revised; specifically, the concept of distance does not seem to be valid
for a system of entangled particles. As has been emphasized, one way to
demonstrate Bell's inequality is to assume ``that there is a well-defined
space of complete states [$\lambda $] of the particle pair and a
well-defined probability distribution [$\rho (\lambda )$] over this space
when an experimental procedure for specifying an ensemble of pairs is
given.''$^{\mathbf{11}}$ Although this may sound like a reasonable
assumption, it can be considered an unnecessarily restrictive one, and
arguments for dispensing with it have been presented, but they have used
counterfactuals, thereby raising some serious criticisms.$^{\mathbf{9,11}}$
Counterfactual reasoning is based on the assumption that we could have
performed another measurement on one of the particles of an entangled pair
(instead of the measurement actually performed, for example) without
changing the outcome of the measurement performed on the other, distant
particle (locality assumption). But it can be argued that in this case we
would be in another Universe, which invalidates the counterfactual-local
reasoning. In particular, assuming that the present is rigidly determined by
the past, two physical events, even separated by a space-like interval, can
be interconnected in such a way that we cannot change one without changing
the other. The introduction of free will, although very reasonable for some,
only makes the argument more disputable. Many different versions of Bell's
theorem have been presented, but they either introduce $\rho (\lambda )$ or
use counterfactuals. Therefore, there seems to be no compelling theoretical
evidence that entanglement implies that quantum mechanics is inherently
nonlocal. However, as we will see, to derive Bell's theorem we do not need
to assume that there is a well-defined $\rho \left( \lambda \right) $ or,
alternatively, that counterfactual reasoning is valid.

%\newpage\

\textbf{IV.\ THE WIGNER-D'ESPAGNAT\ INEQUALITY}

\bigskip\ Our purpose is to answer the question: can a local \textit{theory }%
mimic the \textit{predictions }of quantum mechanics? In other words, we are
interested in situations represented by $(1)$. This expression shows us, in
an ideal situation, what correlations must be observed \textit{if }the
polarizations are measured. That is, we are considering ``latent''
probabilities, so to speak. Actually, there may be no polarizers, that is no
experiment to determine the polarization of the photons. What we want to
know is whether a local theory may have states with the same \textit{latency
}as the states represented by $(1)$. Initially, let us imagine\textit{\ }a
situation in which two-channel polarizers I and II, on which $\nu _1$ and $%
\nu _2$ impinge, respectively, have the same orientation. According to $(1)$%
, whenever $\nu _1$ is transmitted (reflected), $\nu _2$ must also be
transmitted (reflected). Therefore, assuming locality (i.e., what happens to
$\nu _1$ cannot affect $\nu _2$, and vice versa), whether $\nu _1$ and $\nu
_2$ will be transmitted or reflected is already determined before they
impinge on the polarizers; otherwise, we could have a situation in which one
photon is transmitted and the other is reflected. In other words, perfect
correlations \textit{and }locality imply a strong form of determinism.
Assuming that the source has no information about the orientations of the
polarizers, each photon pair has to be emitted with ``instructions,'' so to
speak, for all possible orientations. For instance, transmission in case of
orientation $\mathbf{a}$, reflection in case of orientation $\mathbf{b}$,
and so on. Or, put another way, the outcome of a potential experiment is
determined by the photon's hidden variable state and the polarizer
orientation, and nothing else. As we will see, it is impossible to mimic the
predictions of quantum mechanics in this case.

It is important to mention that in 1982 Itamar Pitowsky published a paper in
which the EPR-Bell ``paradox'' was supposedly solved.$^{\mathbf{12}}$ His
point was that the derivation of Bell's inequality was valid only when a
well-defined $\rho (\lambda )$ could be introduced, which was not the case
for his model. But, as shown by Alan Macdonald,$^{\mathbf{13}}$ in the
particular case of the Pitowsky model, there is another and very simple way
to obtain these inequalities. Actually, a similar derivation had already
been presented by Bernard D'Espagnat in his article on quantum theory and
realism in \textit{Scientific American},$^{\mathbf{14}}$ and before that by
Wigner.$^{\mathbf{15}}$ It is perhaps the simplest and most satisfactory
proof of Bell's inequality, but seldom presented or mentioned; probably
because it is only valid for perfect correlations. Although it is already
very simple, it still can be simplified even further, as we will see.

Let us assume that $N_0$ pairs of photons are emitted. Now let $N\left(
a,b,c\right) $ $\left[ N\left( a_{\perp },b,c\right) \right] $ represent the
number of photon pairs in which the photons are ``prepared'' to be
transmitted when impinging on a polarizer oriented parallel to $\mathbf{b}$
or $\mathbf{c}$, and to be transmitted [reflected] if oriented parallel to $%
\mathbf{a}$. Then, we must have
\begin{equation}
N\left( a,b,c\right) +N\left( a_{\perp },b,c\right) =N\left(
b,c\right) , \label{2}
\end{equation}
where $N(b,c)$ represents the number of photon pairs prepared to be
transmitted when impinging on a polarizer oriented parallel to $\mathbf{b}$
or $\mathbf{c}$. We also must have
\begin{equation}
N(a,c)\geq N(a,b,c)  \label{3}
\end{equation}
and
\begin{equation}
N(a_{\perp },b)\geq N(a_{\perp },b,c).  \label{4}
\end{equation}
From $(2)$, $(3)$ and $(4)$ we obtain
\begin{equation}
N(a,c)+N(a_{\perp },b)\geq N(b,c),  \label{5}
\end{equation}
which is the Wigner-D'Espagnat inequality.$^{\mathbf{16}}$

According to quantum mechanics, we must have $N(a,c)=(N_0/2)\cos ^2(a,c)$, $%
N\left( a_{\perp },b\right) =(N_0/2)\sin ^2\left( a,b\right) $, and $%
N(b,c)=(N_0/2)\cos ^2(b,c)$. Thus, choosing $\left( a,b\right) =\left(
b,c\right) =30^0$, and $\left( a,c\right) =60^0$, we obtain $0.5\geq 0.75$,
violating inequality $(5)$.

\bigskip\

\textbf{V. DISCUSSION}

\bigskip\ As our discussion based on time-like events and entangled
particles has made evident, quantum mechanics is intrinsically a nonlocal
theory, and this conclusion is equally valid for space-like events.
Measuring the polarization state of one of the photons of an entangled
state, represented by $(1)$, instantaneously forces the other, distant
photon into a well-defined polarization state. Paradoxical as it may sound,
this seems to be true independently of the Lorentz frame used to describe
the events (it is important to remember, however, that special relativity is
not necessarily incompatible with the idea of a preferred frame).$^{\mathbf{%
17}}$ In other words, if quantum mechanics is complete, it must be nonlocal.
Since this seems to go against the spirit of special relativity, it is
important to investigate the possibility of quantum mechanics being only a
manifestation of a deeper local theory. As we have seen, perfect
correlations, together with locality, leads to determinism, and determinism
leads to the violation of a Bell inequality. Many derivations of Bell's
inequalities are based on the assumption that there is a well-defined space
of complete states $\lambda $ of the particle pair and a well-defined
probability distribution $\rho (\lambda )$ over this space. Since the
violation of these inequalities might be considered only one proof that this
assumption is false (we might have an infinite and non denumerable number of
hidden variables, for example), it is important to show that it is
unnecessary. As has been emphasized, no counterfactual definiteness needs to
be introduced for this; we only need determinism. Although counterfactual
definiteness presupposes determinism, the converse is not necessarily true.
Strictly speaking, some kinds of determinism---as in the present paper, for
example, in which the outcome of the experiment depends only on the photon
hidden variable state and the orientation of the polarizer---may imply a
sort of virtual contrafactualness, valid in the realm of imagined
experiments, but as a consequence and not as a basic assumption. Actually,
counterfactual definiteness was introduced in an attempt to avoid the use of
hidden variable states.$^{\mathbf{18}}$

A delicate point related to Bell's inequalities involves realism. It was
implicit in our assumption of hidden variable states. It is evident that if
realism is abandoned, it becomes difficult to explain the predicted
correlations assuming locality. Therefore, abandoning realism does not solve
the EPR puzzle. Of course, solipsism is a logical alternative but very
unsatisfactory as a predictive tool and difficult to maintain consistently
in real life. It seems that physics has little to contribute to this
longstanding philosophical debate.

Another delicate point is related to the use of a free-will assumption to
derive Bell's inequalities. According to Bell, ``In the analysis [of EPR
experiments] it is assumed that free will is genuine, and as a result of
that one finds that the intervention of the experimenter at one point has to
have consequences at a remote point, in a way that influences restricted by
the finite velocity of light would not permit. If the experimenter is not
free to make this intervention, if that is also determined in advance, the
difficulty disappears.''$^{\mathbf{19}}$ As has become evident from our
discussion, abandoning free will---which plays no role in our
demonstration--- is not a solution to EPR paradox.

In conclusion:

We have reexamined Bell's theorem using a different approach and trying to
answer the question: Can a local \textit{theory }mimic the \textit{%
predictions }of quantum mechanics? We have assumed that no information about
the orientation of the polarizers is contained in the state of the emitted
pair of photons. In other words, Nature is governed by physical laws;
nothing that might sound like a sort of Big Conspiracy exists. Actually,
there may be no polarizers, that is, no experiment to determine the
polarization of the photons. We only know what correlations must be observed
\textit{if }the polarizations are measured. This is in agreement with
expression $(1)$, which only expresses potentialities. Strictly speaking, we
are not discussing whether Nature is nonlocal or not---although experimental
evidence strongly suggests it is$^{\mathbf{20}}$---but whether quantum
mechanics is nonlocal or not. In our demonstration, neither the assumption
that there is a well-defined space of complete states of the particle pair
and a well-defined probability distribution over this space nor the use of
counterfactuals is needed. This makes the conclusion that quantum mechanics
is intrinsically nonlocal almost inescapable.

\bigskip\ \ \

\noindent \textbf{REFERENCES AND NOTES}

\bigskip\

\noindent $^{\mathbf{1}}$E. Schr\"{o}dinger, ``Discussion of probability
relations between separated systems,'' Proc. Cambridge Phil. Soc. \textbf{31}%
, 555-563 (1935).

\noindent $^{\mathbf{2}}$A. Einstein. B. Podolsky, and N. Rosen, ``Can
quantum mechanical description of physical reality be considered
complete?,'' Phys. Rev. \textbf{47}, 777-780 (1935), reprinted in \textit{%
Quantum Theory and Measurement}, edited by J. A. Wheeler and W. H. Zurek
(Princeton University Press, Princeton, 1983).

\noindent $^{\mathbf{3}}$J. S. Bell, ``On the Einstein-Podolsky-Rosen
paradox,'' Physics \textbf{1}, 195-200 (1964), reprinted in \textit{%
Speakable and unspeakable in quantum mechanics} (Cambridge University Press,
Cambridge, 1987) and in \textit{Quantum Theory and Measurement}, edited by
J. A. Wheeler and W. H. Zurek (Princeton University Press, Princeton, 1983).

\noindent $^{\mathbf{4}}$J. F. Clauser, M. A. Horne, A. Shimony, and R. A.
Holt, ``Proposed experiment to test local hidden-variables theories,'' Phys.
Rev. Lett. \textbf{23}, 880-884 (1969), reprinted in \textit{Quantum Theory
and Measurement}, edited by J. A. Wheeler and W. H. Zurek (Princeton
University Press, Princeton, 1983), J. S. Bell, ``Introduction to the
hidden-variable question,'' in \textit{Foundations of Quantum Mechanics},
edited by B. D'Espagnat (Academic, New York, 1971), reprinted in \textit{%
Speakable and unspeakable in quantum mechanics} (Cambridge University Press,
Cambridge, 1987); J. F. Clauser and M. A. Horne, ``Experimental Consequences
of Objective Local Theories,'' Phys. Rev. D \textbf{10}, 526-535 (1974); L.
C. Ryff, ``Bell and Greenberger, Horne, and Zeilinger theorems revisited,''
Am. J. Phys. \textbf{65}, 1197-1199 (1997).

\noindent $^{\mathbf{5}}$D. M. Greenberger, M. Horne, and A. Zeilinger,
``Going Beyond Bell's Theorem,'' in \textit{Bell's Theorem, Quantum Theory,
and Conceptions of the Universe}, edited by M. Kafatos (Kluwer Academic,
Dordrecht, 1989); D. M. Greenberger, M. A. Horne, A. Shimony, and A.
Zeilinger, ``Bell's theorem without inequalities,'' Am. J. Phys. \textbf{58}%
, 1131-1143 (1990); L. Hardy, ``Nonlocality for two particles without
inequalities for almost all entangled states,'' Phys. Rev. Lett. \textbf{71}%
, 1665-1668 (1993).

\noindent $^{\mathbf{6}}$S. Fasel, N. Gisin, G. Ribordy, and H. Zbinden,
``Quantum key distribution over 30 km of standard fiber using energy-time
entangled photon pairs: a comparison of two chromatic dispersion reduction
methods,'' quant-ph/0403144 (2004); I. Marcikic, H. de Riedmatten, W.
Tittel, H. Zbinden, M. Legr\'{e}, and N. Gisin, ``Distribution of time-bin
entangled qubits over 50 km of optical fiber,'' quant-ph/0404124 (2004);
C-Z. Peng, T. Yang, X-H. Bao, J-Zang, X-M. Jin, F-Y. Feng, B. Yang, J. Yang,
J. Yin, Q. Zhang, N. Li, B-L. Tian, and J-W. Pan,``Experimental Free-Space
Distribution of Entangled Photon Pairs over a Noisy Ground Atmosphere of
13km,'' quant-ph/0412218 (2004).

\noindent $^{\mathbf{7}}$N. Gisin, G. Ribordy, W. Tittel, and H. Zbinden,
``Quantum cryptography,'' quant-ph/0101098 (2001); M. Aspelmeyer, T.
Jennewein, A. Zeilinger, M. Pfennigbauer, and W. Leeb, ``Long-Distance
Quantum Communication with Entangled Photons using Satellites,''
quant-ph/0305105 (2003); C. Elliot, D. Pearson, and G. Troxel, ``Quantum
Cryptography in Practice,'' quant-ph/0307049 (2003); R. Kaltenbaek, M.
Aspelmeyer, T. Jennewein, C. Brukner, A. Zeilinger, M. Pfennigbauer, and W.
Leeb, ``Proof-of-Concept Experiments for Quantum Physics in Space,''
quant-ph/0308174 (2003); R. Thew, A. Ac\'{\i}n, H. Zbinden, and N. Gisin,
``Experimental Realization of Entangled Qutrits for Quantum Communication,''
quant-ph/0307122 (2004); A. Poppe, A. Fedrizzi, T. Lor\"{u}nser, O.
Maurhardt, R. Ursi, H. B\"{o}hm, M. Peev, M. Suda, T. Jennewein, and A.
Zeilinger, ``Practical Quantum Key Distribution with Polarization Entangled
Photons,'' quant-ph/0404115 (2004); C. Elliot, A. Colvin, D. Pearson, O.
Pikalo, J. Schlafer, and H. Yeh, ``Current status of the DARPA Quantum
Network,'' quant-ph/0503058 (2005).

\noindent $^{\mathbf{8}}$N. Gisin, ``How come the Correlations?,''
quant-ph/0503007 (2005); L. C. Ryff, ``Interference, distinguishability, and
apparent contradiction in an experiment on induced coherence,'' Phys. Rev.
A. \textbf{52}, 2591-2596 (1995); L. C. Ryff, ``The Strange Behavior of
Entangled Photons,'' Found. Phys. Lett. \textbf{10}, 207-220 (1997); L. C.
Ryff, ``Two-photon interference without intrinsic indistinguishability,''
Quantum Semiclass. Opt. \textbf{10}, 409-414 (1998); L. C. Ryff,
``Interaction-Free Which-Path Information and Some of Its Consequences,'' Z.
Naturforsch. \textbf{56a}, 155-159 (2001).

\noindent $^{\mathbf{9}}$A. Shimony, ``An Analysis of Stapp's `A Bell-type
theorem without hidden variables','' quant-ph/0404121 (2004), and references
therein; H. P. Stapp, ``Comments on Shimonys's Analysis,'' quant-ph/0404169
(2004).

\noindent $^{\mathbf{10}}$Interviews with John Bell and David Bohm in
\textit{The Ghost in the Atom, }edited by P. C. W. Davies and J. R. Brown
(Cambridge University Press, Cambridge, 1986). The first quotation is from
page 49, the second from page 125.

\noindent $^{\mathbf{11}}$A. Shimony, ``An exposition of Bell's theorem,''
in \textit{Search For A Naturalistic World View, vol. II }(Cambridge
University Press, Cambridge, 1993), p. 103.

\noindent $^{\mathbf{12}}$I. Pitowsky, ``Resolution of the
Einstein-Podolsky-Rosen and Bell Paradoxes,'' Phys. Rev. Lett. \textbf{48},
1299-1302 (1982).

\noindent $^{\mathbf{13}}$A. Macdonald, ``Comment on `Resolution of the
Einstein-Podolsky-Rosen and Bell Paradoxes','' Phys. Rev. Lett. \textbf{49},
1215 (1982). I thank Prof. Macdonald for kindly calling my attention to his
Comment.

\noindent $^{\mathbf{14}}$B. D'Espagnat, ``The Quantum Theory and Reality,''
Scientific American\textit{\ }\textbf{241}, 128-140 (1979).

\noindent $^{\mathbf{15}}$E. P. Wigner, ``On hidden variables and quantum
mechanical probabilities,'' Am. J. Phys. \textbf{38}, 1005-1009 (1970); E.
P. Wigner, ``Interpretation of Quantum Mechanics,'' lectures given in the
Physics Department of Princeton University during 1976, revised and printed
in \textit{Quantum Theory and Measurement}, edited by J. A. Wheeler and W.
H. Zurek (Princeton University Press, Princeton, 1983).

\noindent $^{\mathbf{16}}$J. S. Bell, ``Bertlman's socks and the nature of
reality,'' Journal de Physique \textbf{42}, 41-61 (1981), reprinted in
\textit{Speakable and unspeakable in quantum mechanics} (Cambridge
University Press, Cambridge, 1987).

\noindent $^{\mathbf{17}}$J. S. Bell, ``How to teach special relativity,''
Progress in Scientific Culture, Vol. 1, No 2, summer 1976, reprinted in
\textit{Speakable and unspeakable in quantum mechanics (}Cambridge
University Press, Cambridge, 1987\textit{).}

\noindent $^{\mathbf{18}}$J. F. Clauser and A. Shimony, ``Bell's theorem:
experimental tests and implications,'' Rep. Prog. Phys. \textbf{41},
1881-1927 (1978).

\noindent $^{\mathbf{19}}$Interview with John Bell in \textit{The Ghost in
the Atom, }edited by P. C. W. Davies and J. R. Brown (Cambridge University
Press, Cambridge, 1986). The quotation is from page 47.

\noindent $^{\mathbf{20}}$S. J. Freedman and J.F. Clauser, ``Experimental
test of local hidden-variable theories,'' Phys. Rev. Lett. \textbf{28},
938-941 (1972), reprinted in \textit{Quantum Theory and Measurement}, edited
by J. A. Wheeler and W. H. Zurek (Princeton University Press, Princeton,
1983); A. Aspect, P. Grangier, and G. Roger, ``Experimental realization of
Einstein-Podolsky-Rosen-Bohm gedankenexperiment: a new violation of Bell's
inequalities,'' Phys. Rev. Lett. \textbf{49}, 91-94 (1982); A. Aspect, J
Dalibard, and G. Roger, ``Experimental test of Bell's inequalities using
time-varying analyzers,'' Phys. Rev. Lett. \textbf{49}, 1804-1807 (1982); Y.
H. Shih and C. Alley, ``New type of Einstein-Podolsky-Rosen- Bohm experiment
using pairs of light quanta produced by optical parametric down-conversion,
Phys. Rev. Lett. \textbf{61}, 2921-2924 (1988); G. Weihs, T. Jennewein, C.
Simon, H. Weinfurter, and A. Zeilinger, ``Violation of Bell's inequality
under strict Einstein locality conditions,'' Phys. Rev. Lett. \textbf{81},
5039-5043 (1998).

\end{document}